\documentclass[fleqn,twoside,twocolumn,nofootinbib]{revtex4} 
\usepackage[sec]{ujp} 
\def\y{\tilde y}
\def\x{\tilde x}
\def\*{\noindent}

\begin{document}
\title[CAUSTIC CROSSING EVENTS AND SOURCE MODELS]
{CAUSTIC CROSSING EVENTS AND SOURCE MODELS\\ IN GRAVITATIONAL LENS SYSTEMS}%
\author{A.N.~Alexandrov}
\author{V.M.~Sliusar}
\author{V.I.~Zhdanov}%
\affiliation{Taras Shevchenko National University of Kyiv}
\address{2, Academician Glushkov Prosp., Kyiv 03022, Ukraine}
\email{alex@observ.univ.kiev.ua; vitaliy.slyusar@gmail.com; zhdanov@observ.univ.kiev.ua}

\udk{524.8}  \razd{\secxi}

\setcounter{page}{389}%
\maketitle

\begin{abstract}
High amplification events (HAEs) are common phenomena in
extragalactic gravitational lens systems (GLSs), where the multiple
images of a distant quasar are observed through a foreground galaxy.
There is a considerable brightness magnification in one of the
quasar images during HAE. Grieger, Kayser, and Refsdal (1988)
proposed to use HAEs to study the central regions of quasars in
GLSs. In this paper, we consider some problems concerning the
identification of different source types on the basis of the HAE
observations. We compare the results of light curve simulations to
estimate a feasibility to distinguish different source models in
GLSs. Analytic approximation methods yielding solutions of the lens
equation in a vicinity of fold caustic crossing events are
presented. The results are used to obtain amplification factors,
which the higher-order corrections for the Gaussian, power-law, and
limb-darkening models of a source take into account.
\end{abstract}

\section{Introduction}
\label{Introduction}

Gravitational lens systems can be viewed as natural telescopes
that provide a valuable information about remote objects. In an
extragalactic gravitational lens system (GLS), a distant quasar is
observed through a foreground galaxy. The gravitational field of
the galaxy can bend light rays from the source sufficiently
so that there are multiple light rays that reach an
observer. The observer sees an image in the direction of each ray,
so that the source appears multiply imaged.

Light rays from a quasar pass through the lensing galaxy in
different regions. Local variations of gravitational fields in
these regions, which are mainly due to a relative motion of the
lensing galaxy and the source, lead to considerable brightness
variations in each image (gravitational microlensing), which can
be detected even by modest telescopes. The typical time scales of
these microlensing processes may vary from weeks to months. A
comparison of the independent brightness variations in different
images provides a valuable information about the lensing galaxy
and about the source structure \citep{schneider_92,Wambsganss_06}.
One of the most important applications of the microlensing
concerns a unique possibility to study a fine structure of the
central quasar region that cannot be resolved in another way with
modern observational techniques. This is important because the
quasars in GLSs have high redshifts. Therefore, when studying the
quasars, we learn something about a corresponding early epoch.

HAEs -- considerable brightness
magnifications in one of the images of a quasar -- are common
phenomena in extragalactic GLSs. Grieger, Kayser and Refsdal \cite
*{grieger_88}  proposed to use HAEs to study the central regions of
the quasars in GLSs. Typically corresponding variations of the
brightness in a neighborhood of the HAE can be approximately described
by a formula containing a few fitting parameters. This makes it
possible to estimate some GLS parameters, in particular, the
source size \citep{grieger_88}. For example, in case of the
well-known GLS Q2237+0305 ``Einstein Cross'' \citep{Huchra_85}, several HAEs were
observed \citep{wozniak_00, alcald_02, udalski_2006}, and the
estimates of the source size have been obtained for different
source  models
\citep{wyithe_99,wyithe_00b,wyithe_00a,yonehara_01,shalyapin_01,shalyapin_02,Bogdanov_02}.
Almost all HAEs in the Q2237+0305 GLS are attributed to the
intersection of a fold caustic in the source plane (see, e.g.,
\citep{gil_06}). In view of the increasing accuracy of photometric
observations, a possibility to distinguish different source models
is also discussed (see, e.g., \citep{goico_01,Mortonson_05}).

Here, we study some problems dealing mainly with the investigation
of the source structure in an extragalactic GLS using the light
curves of the source images. We discuss the results of simulations
of light curves for different microlensed source models and
present some analytic approximations yielding solutions of the
lens equation in a vicinity of the fold caustic crossing events.
The results are used to obtain amplification factors for some
source models that take the ``post-linear'' corrections into
account.

 Note that, in reality, we have, indeed, a unique
light curve for each image of the quasar in a GLS. The high amplification events are not too frequent, and
it may take a considerable time to accumulate a sufficient
statistics and even to wait a repetition of the HAE. Thus,
the main attention is paid to the determination of source properties
from a single light curve compared with the light curves of the
other images. On the other hand, since the work by Kochanek
\citep{Kochanek_04} followed by a number of works
\citep{Mortonson_05,gil_06,vakulik_07,anguita_08, Poindexter_08,
Poindexter_10a, Poindexter_10b}, numerous statistical methods have
been developed to process complete light curves. These methods treat
the whole available light curves from all the images, not only HAEs.
There is an enormous number of the degrees of freedom which prohibit
the simultaneous determination of {\textit {all}} the parameters of a GLS
(microlens positions, their masses, and the source parameters). But
the observational data restrict possible realizations of the microlens
masses and the positions, so one may estimate the conditional
probability of certain parameter values. Such approach is very
attractive because it allows one to use the whole aggregate of
observational data on image light curves. However, this involves a
large number of realizations of the microlensing field and,
respectively, a considerable computer time.\looseness=1

On the other hand, the source structure manifests itself only in
HAEs; far from the caustic, the source looks like a point one, and
all the information about its structure is lost. If we restrict
ourselves to a HAE neighborhood, then we use the most general model
concerning a microlensing field described by a small collection of
the Taylor expansion coefficients in the lens mapping.

The structure of this paper is as follows. Section~\ref{s1}
contains some basics of the gravitational lensing, a short review
of the problems concerning the determination of a source model from
HAE observations, and a list of typical models. In
Section~\ref{simulations}, we present the results of simulations
for these source models, in particular, the differences of light
curves. In Section~\ref{s2}, we outline approximation methods to
investigate solutions of the lens equation used to derive the
amplification coefficient of a point source. The result is used to
obtain amplifications of an extended source near the fold caustic
for some source models. Section~\ref{discussion} contains the
discussion of the results.

\section{General Relations and Notations}
\label{s1}

\subsection{The lens equation}
\label{s1a}

First, we recall some general notions concerning the gravitational
lensing that can be found, e.g., in book \cite{schneider_92}.
The normalized lens equation has the form
\begin{equation}
\label{eq1}
{\rm {\mathbf y}} = {\rm {\mathbf x}} - \nabla \Phi
\left( {\rm {\mathbf x}} \right),
\end{equation}

\noindent where $\Phi \left( {\rm {\mathbf x}} \right)$ is the
lens potential. This equation relates every point ${\rm {\mathbf
x}}=(x_1,x_2)$ of the image plane to the point ${\rm {\mathbf
y}}=(y_1,y_2)$  of the source plane. In the general case, there
are several solutions ${\mathbf X}_{\left( i \right)} \left(
{\mathbf y} \right)$ of the lens equation (\ref{eq1}) that
represent images of one point source at $\mathbf y$; we denote the
solution number by the index in parentheses.

If there is no continuous matter on the line of sight, the potential
must be a harmonic function: $\Delta \Phi = 0$. Below, we assume
that this condition is fulfilled in a neighborhood of the critical
point. We note, however, that if the continuous matter density is
supposedly constant during a HAE, this can be taken into account by
a suitable renormalization of the variables.

The amplification of a separate image of the point source is
\begin{equation}
\label{eq2}
K_{(i)} \left( {\rm {\mathbf y}} \right) = 1
\mathord{\left/ {\vphantom {1 {\left| {J\left( {{\rm {\mathbf
X}}_{\left( i \right)} \left( {\rm {\mathbf y}} \right)} \right)}
\right|}}} \right. \kern-\nulldelimiterspace} {\left| {J\left(
{{\rm {\mathbf X}}_{\left( i \right)} \left( {\rm {\mathbf y}} \right)}
\right)} \right|},
\end{equation}

\noindent where $J\left( {\rm {\mathbf x}} \right) \equiv \left|
{{D\left( {\rm {\mathbf y}} \right)} \mathord{\left/ {\vphantom
{{D\left( {\rm {\mathbf y}} \right)} {D\left( {\rm {\mathbf x}}
\right)}}} \right. \kern-\nulldelimiterspace} {D\left( {\rm {\mathbf
x}} \right)}} \right|$ is the Jacobian of the lens mapping. In the
microlensing processes, microimages cannot be observed separately;
therefore, we need the total amplification that is a sum of the
amplification factors of all the images.

The critical curves of the lens mapping~(\ref{eq1}) are determined
by the equation $J\left( {\rm {\mathbf x}} \right) = 0$ and are
mapped onto the caustic in the source plane. The stable critical
points of a two-dimensional mapping can be folds and cusps only,
the folds being more probable in a HAE. In this paper, we confine
ourselves to the consideration of a fold caustic. When the point
source approaches the fold caustic from its convex side, two of
its images approach the critical curve, and their amplification
tends to infinity. They disappear when the source crosses the
caustic. These two images are called critical.

\subsection{Problems}
\label{bri}

Let $I({\mathbf y})$ be the initial surface brightness distribution
of an extended source.  If the source center is located at the
point ${\mathbf{Y}}=(Y_1,Y_2)$ in the source plane, then the total
microlensed flux from the source is
\begin{equation}
\label{flux_extended1}
 F( {\mathbf Y})= \int\!\!\!\int I ( { {\mathbf y}( {\mathbf x} )-{\mathbf Y}})\, dx_1
 dx_2 ,
\end{equation}
\noindent ${\mathbf x}=(x_1,x_2)$. The result of using
Eq.~(\ref{flux_extended1}) is obviously equivalent to the result
of the well-known ray-tracing method \citep{schneider_92}
(when the pixel sizes tend to zero).

An equivalent representation of this formula is
\begin{equation}
\label{flux_extended}
 F( {\mathbf Y})=  \int\!\!\!\int K( {\rm {\mathbf y} })
 I ( {\rm {\mathbf y}-{\mathbf Y}})\,dy_1 dy_2,
\end{equation}

\noindent where the point source amplification $K( {\mathbf y} ) =
\sum\limits_i {K_i (\mathbf y)} $ is the sum of amplifications of
all the images.

Near the caustic, one can approximate $K( {\mathbf y} )=K_0 + K_{\rm
cr} ( {\mathbf y})$, where $K_0 $ is the amplification of all
noncritical images that is supposed to be constant during HAE, and
$K_{\rm cr}$ is the amplification of the critical images. Due to a
relative motion of the lensing galaxy  and the source (quasar), the
flux is a function of time representing the light curve of some
quasar image in the GLS. In the lensing galaxy rest frame, the
quasar motion may be considered as a straight-line motion with a
sufficient accuracy. Here, we are mostly interested in the caustic
crossing events when the quasar intersects the caustic in the source
plane leading to a considerable enhancement of the image brightness.
The light curve yields information about the source structure and
about the gravitational field of the lensing galaxy. Particularly,
the form of a light curve near a HAE depends on the function
$I({\rm {\mathbf y}})$, and the question is whether it is possible
to use this dependence in order to identify this function.

To illustrate the typical problems arising in the conventional treatment
of a HAE, we consider an approximate relation for the
amplification coefficient $K_{\rm cr}({\mathbf{y}})$ of a point
source located at the point ${\mathbf{y}}=(y_1,y_2)$ of the source
plane which is located near the fold. In the appropriate coordinate
system, the amplification of a point source can be approximated
 as $K_{\rm cr}(y_1,y_2)=A_0 (y_2)^{-1/2} \Theta(y_2)$ \citep{schneider_92}, where
$y_2$ is the distance between the point source and the caustic, and
$\Theta(y_2)$ is the Heaviside step function. The observed
radiation flux from the extended source image during the HAE is then
obtained from Eq.(\ref{flux_extended}) yielding an integral
equation for the one-dimensional luminosity profile $f(y_2)=\int
I(y_1,y_2) dy_1 $:
\begin{equation}
\label{flux_extended linca}
 F(t) = C_1+A_0  \int \Theta(y_2) (y_2)^{-1/2}  f( y_2-Y_2(t)) dy_2,
\end{equation}
\noindent ${\mathbf Y}(t)=(Y_1(t),Y_2(t))$ is the source center
trajectory which can be written as a linear function of time;
$F(t)$ is known from observations, $C_1$ describes a contribution
of noncritical images and can be considered to be constant during
the HAE. Thus, we have an equation for $f(y)$.

The main problems concerning with this equation are as follows.

(i) Equation (\ref{flux_extended linca}) gives us nothing about
the whole function $I(y_1,y_2)$ we are interested in, unless some
suppositions about the form of the source are made, e.g., the
supposition about circular symmetry. Also, even for known $F(t)$
and $Y_2(t)$,  Eq. (\ref{flux_extended linca}) presents a kind of
the ill-posed mathematical problems: small variations of input
data can lead to a considerable change of the solution. A standard
way to relax this difficulty involves additional restrictions on
$f(y)$ and/or the use of some explicit models for the brightness
distribution $I({\bf y})$ containing a small number of free
parameters. Some of these models are considered below in the next
subsection.

(ii) It is clear that a real brightness profile of the central
quasar region is quite different from the simplified brightness
distributions of the following section that can be considered rather
as some reference models. However, in view of the present-day accuracy
of observations, sometimes it can be difficult to distinguish even
these simple source models using a HAE. For example, the authors of
\cite{Mortonson_05} argue that the light curves from a accretion disk can
be well fitted with any brightness profile (Gaussian, uniform, {\it
etc.}) of an appropriate source size. On the other hand, a number
of authors \citep{shalyapin_01,shalyapin_02,Bogdanov_02,
goico_01,Kochanek_04, Mortonson_05,gil_06, vakulik_07,anguita_08}
discussed the delicate questions concerning the determination a fine quasar
structure from HAEs. For example, the authors of \cite{goico_01}
wrote that the GLITP data \citep{alcald_02} on Q2237+0305 admit
only accretion disc models (see also \citep{gil_06, anguita_08}).
Obviously, the presence of an accretion disk in the central region of
a quasar is beyond any doubts, as well as the fact that the real
quasar core can be quite different from its
simplified models in question. The question is how to prove \textit{a posteriori}
the existence of the accretion disk on a basis of
available observational data.\looseness=1

One must also have in mind that, without using an additional
information (besides HAEs), we cannot determine even the source size,
because we do not know the value and the direction of the source
velocity with respect to the caustic. Also, different ellipticities
and orientations of the source may lead to different forms of the light
curves during a HAE.

(iii) The kernel $K_{\rm cr}$ of the integral equation
(\ref{flux_extended linca}) is a result of the so-called linear
caustic approximation of the lens equation. This will work correctly
only in the case where the source size is much smaller than the caustic
curvature radius. Below, we consider corrections to $K_{\rm cr}$ that
arise in the post-linear approximation. However, these approximations
also require the source to be sufficiently small. This requirement
can be violated in case of a complicated caustic network or in the
presence of a population of small microlenses (cf. planetary
masses).

\subsection{The extended source models}
\label{extended}

Below, we list the simplest most commonly used brightness
distributions of the source in a GLS; without loss of generality, they
are chosen to be normalized to 1:
\[
\int\!\!\!\int  I ( {\rm {\mathbf y}})\,dy_1 dy_2=1.
\]

\noindent To compare different models of brightness distribution,
we have to use the same parameter that characterizes the size of an
object. The most general is the r.m.s. size $R_{\rm rms}$:
\begin{equation}
\label{Rrms} R_{\rm rms}^2=\int\!\!\!\int {\mathbf y}^2 I ( {\rm
{\mathbf y}})\,dy_1 dy_2.
\end{equation}
\noindent For a slowly decreasing brightness profile (e.g., $I (\rm
{\mathbf y})\sim |{\mathbf y}|^{-\alpha},\,\,\, \alpha \le 4 $ )
the r.m.s. size diverges. In the case of the circularly symmetric
sources, the half-brightness radius $R_{1/2}$ is also widely used;
it is defined by the relation
\begin{equation}
\label{Rhalf} 2\pi \int\limits_{0}^{R_{1/2}} \! I(r)r \, \mathrm{d}r
= \pi \int\limits_{0}^{\infty} \! I(r)r \, \mathrm{d}r .
\end{equation}

In the case of the Gaussian source model,
\begin{equation}
\label{Gauss} I_G(r)=\frac{1}{\pi R^2}
\exp\left[-\left(\frac{r}{R}\right)^2\right],
\end{equation}
where $R=R_{\rm rms}$, $ R_{1/2} = R \sqrt{\ln(2)}$, and $R$ stands for
a size parameter.

The limb-darkening (LD) model (see, e.g., \citep{dominik}) yields
\begin{equation}
\label{C2} I_{\rm LD} ( r) = \frac{q + 1}{\pi R^2} \, \Xi ( {r}/R;q
),
\end{equation}
where
\[{ \Xi ( \xi;q ) = \Theta (1 - \xi ^2)(1 - \xi ^2)^q},\]
\noindent    and $R^2 = \left( {q + 2} \right)R_{\rm rms}^2 $. Here,
we assume $q > 0$. The half-brightness radius  is
$R_{1/2}=R\sqrt{1-(1/2)^{1/(q+1)}}$.

Models (\ref{C2}) and (\ref{Gauss}) describe a class of
compact sources with a fast decrease in the brightness. On the contrary,
the power-law models \citep{shalyapin_01,shalyapin_02} describe a
slower decrease at large $r$:
\begin{equation} \label{C1}
 I_{\rm PL} ( r ) = \frac{p - 1}{\pi R^2}\left[ 1 +
 r^2 /R^2 \right]^{ - p},
\end{equation}
\* where $p > 1$ is the exponent, and $R$ is related to the
r.m.s. radius $R_{\rm rms}$ as $R^2 = \left( {p - 2} \right)R_{\rm
rms}^2 $. Model (\ref{C1}) is an alternative to (\ref{C2}). The
half-brightness radius of the source within this model is
$R_{1/2}=R\sqrt{2^{1/(p-1)} - 1}$. For fixed $R_{\rm rms} $ and $p
\to \infty,$ the brightness distribution (\ref{C1}) tends to the
Gaussian one. For small $p,$ we have a ``long-range'' distribution;
 $R_{\rm rms}$ diverges for $p\le 2$.

Linear combinations of different distributions with different
parameters yield rather a wide class of symmetric source models.

A more complicated profile is presented by the Shakura--Sunyaev model (AD) of
accretion disk \citep{schakura_73}
\begin{equation}
\label{AD} I_{\rm AD}(r)=\frac{3R\theta(r-R)}{2\pi
r^3}\left[1-\sqrt{\frac{r}{R}}\right].
\end{equation}
 For the AD
model, the half-brightness radius is $R_{1/2}=4R$, $R_{\rm
rms}=\infty$

Another  model  \citep{schakura_73,Kochanek_04} (AD1) can be
defined as
\begin{equation}
\label{AD1} I_{\rm AD1}(r)=\frac{C_{\rm AD1}}{ R^2} \left[\exp
(\rho^{3/4})-1\right]^{-1},\quad \rho=r/R,
\end{equation}
\[
C_{\rm AD1}=3(8 \pi \Gamma(8/3)\zeta(8/3))^{-1}\approx 0.06,
\]
$\zeta(x)$ is the Riemann zeta function, $\Gamma(x)$ is the Gamma function, and
\[
R_{\rm rms}^2=R^2
\frac{\Gamma(16/3)\zeta(16/3)}{\Gamma(8/3)\zeta(8/3)}\approx 21.4
R^2 .
\]
It is also interesting to study a superposition of the LD and AD
 models (LA), where the accretion disk has a boundary
\begin{equation}
\label{LA} I_{\rm LA}(r) =\frac{C(q)   \Xi ( \rho;q )}{R^2
\left[\exp (\rho^{3/4})-1\right]^{-1}},\quad \rho=r/R,
\end{equation}
where $C(q)$ is a normalizing coefficient;  $C(1)= 0.42,\,C(2)=
0.54,\, C(3)= 0.64$.

\section{Simulations of Light Curves}
\label{simulations}

\subsection{Equations of microlensing}
\label{EquationsMicrolensing}

In the case without continuous matter on the line of sight, the
lens equation takes the form
\begin{equation}
\label{lens} {\mathbf y}={\mathbf x}-R_{\rm E}^2\sum_{i = 1}^N
\frac{{\mathbf x}-{\mathbf x_i}}{|{\mathbf x}-{\mathbf x_i}|^2},
\end{equation}
where ${\mathbf x_i}$ are positions of the microlenses in the lens
plane, and $R_{\rm E}$ is the radius of the Einstein ring for one
microlens which is assumed to be the same,  $R_{\rm E}=1$, for all
microlenses.

Here, we present the results of straightforward calculations of the
microlensed flux (\ref{flux_extended1}) to obtain the light curves
for different realizations of the microlens positions. All
calculations were performed for the microlensing optical depth
$\sigma=0.3$. The total number of microlenses was 1470.

\begin{figure}
\includegraphics[width=\column]{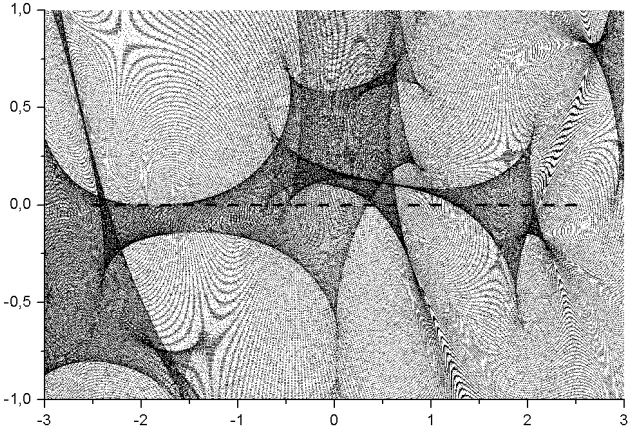}
\vskip-3mm\caption{Magnification pattern with the trajectory of the
source in the source plane (coordinates are measured in $R_{\rm E}$
units). The source moves from left to right along the straight line
with uniform speed}
\label{caustics}
\end{figure}

\begin{table}[b]
\noindent\caption{Parameters of simulation}\vskip3mm\tabcolsep22.2pt
\label{parameters_table} \noindent{\footnotesize
\begin{tabular}{l c}
 \hline%
 \multicolumn{1}{c}{\rule{0pt}{9pt}Parameter}%
 & \multicolumn{1}{|c}{Value}\\%
\hline%
\rule{0pt}{9pt}Number of pixels & $1.23\times10^6$ \\
Pixel size & 0.01 $R_{\rm E}$\\%
Source trajectory length & 2 $R_{\rm E}$; 5 $R_{\rm E}$\\
Radius of field & 70 $R_{\rm E}$\\
Microlensing optical depth ($\sigma$) & 0.3\\
Source speed ($V$) & 1\\
Time discretization ($\delta t$) & 0.01 $R_{\rm E}/V$\\
\hline
\end{tabular}
}
\end{table}

\begin{figure}
\includegraphics[width=\column]{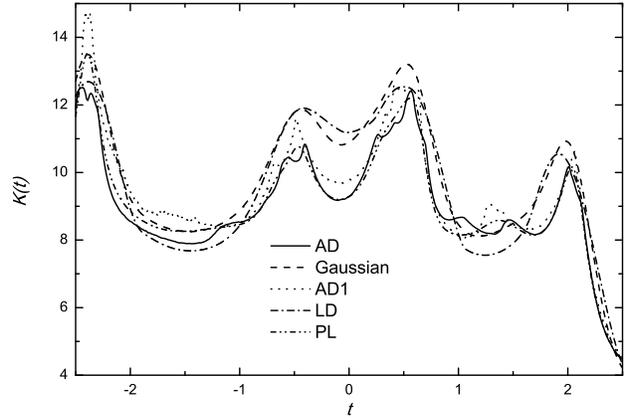}
\vskip-3mm\caption{``Light curves'': the amplification as a function
of time for different source models that correspond to the
magnification pattern in Fig.~\ref{caustics}} \label{lightcurves}
\end{figure}

The microlens positions were chosen in a random way with uniform
distribution over the field.  The trajectory length has been
taken so as to provide the caustic crossings. The size of the
microlens field was chosen large enough to avoid boundary effects.

\subsection{Light curves}
\label{LightCurves}

The form of the light curve of the microlensed source in caustic
crossing events depends on details of the source structure. However,
the point is to determine the most relevant source model with regard
for this form. In this work, we compare the light curves within the
Gaussian, PL, LD, AD, and AD1 source models. The simulations were
performed for the set of 100 realizations of a microlensing field
with the optical depth $\sigma=0.3$ corresponding to parameters of
Q2237+0305 \citep{Schmidt_98,Dai_03, Fedorova_09, Vakulik_06}. All
models have the same half-brightness radius $R_{1/2}.$ The
calculations were performed for the same microlenses fields. The
source speed is $V=1,$ so we can identify the source position as a
function of time $t$. All the microlenses are static. The typical
magnification pattern is shown in Fig.~\ref{caustics}.\looseness=1

First, we present the results of simulations with the same
half-brightness radius $R_{1/2}=0.21$; the power-law exponent was
$p=3/2$ for the ``long range'' PL model; the AD model also corresponds to
this class of the power-law asymptotic dependence with $p=3/2$). For
LD and LA models, we have chosen $q=1$ throughout the paper. From the
``light curves'' in Fig.~\ref{lightcurves}, we observe a significant
difference between the compact (LD and Gaussian) and
``long-range'' models. The long-range character of the latter
reveals itself even at considerable distances from the caustic,
where we expect that the brightness of all sources must have the
same behavior as that of a point source. The differences between
these two groups of models are essentially larger than the
differences within each group (e.g., between the Gaussian and LD models).
This conclusion is confirmed by the results of statistical
considerations over 100 realizations shown in Table
\ref{differences1} for the half-brightness radius $R_{1/2}=0.21$ as an
example.\looseness=1

To compare different models $i$ and $j,$ we used the relative
difference
\begin{equation}
\label{diffs} \eta = 2 \max_t \left(\frac{\left |
K_i(t)-K_j(t)\right |}{K_i(t)+K_j(t)}\right),
\end{equation}
where $K_i(t)$ is the amplification for the $i$-th model along the
trajectory of source's linear movement.

\begin{figure}
\includegraphics[width=\column]{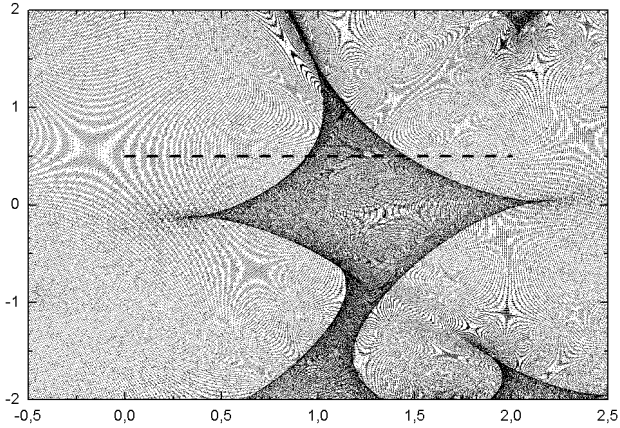}
\caption{Magnification pattern with simple caustic crossing events}
\label{simplecau}
\end{figure}

\begin{table}[b]
 \noindent\caption{Relative difference between models in
HAE}\vskip3mm\tabcolsep16.9pt \label{differences1}
\noindent{\footnotesize
\begin{tabular}{c c c}
 \hline%
 \multicolumn{1}{c}{\rule{0pt}{9pt}$i$-th model}%
 & \multicolumn{1}{|c}{$j$-th model}
 & \multicolumn{1}{|c}{\bf $\eta$ }\\%
\hline%
\rule{0pt}{9pt}{ AD } & { Gaussian } & { $0.074 \pm 0.0012 $ }\\
{ AD  } & { AD1  } & { $0.085 \pm 0.002$ } \\
{ AD  } & { LD } & { $0.091 \pm 0.002$ } \\
{ AD  } & { PL  } & { $0.038 \pm 0.001$ } \\
{ Gaussian } & { AD1 } & { $0.073 \pm 0.0017$ } \\
{ Gaussian } & { LD } & { $0.042 \pm 0.001$ } \\
{ Gaussian } & { PL } & { $0.073 \pm 0.0013$ } \\
{ AD1  } & { LD } & { $0.094 \pm 0.002$ } \\
{ AD1 } & { PL }  &  { $0.052 \pm 0. 0012$}\\
{ LD } & { PL } & { $0.090 \pm 0.002$ } \\
\hline
\end{tabular}
}
\end{table}

\begin{figure}
\includegraphics[width=\column]{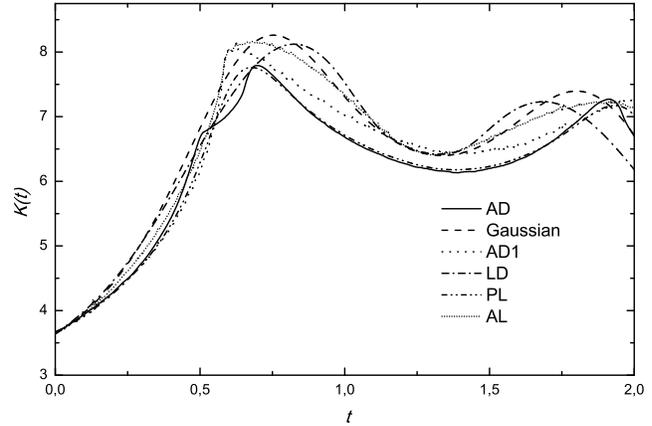}
\caption{Light curves (magnification curves) for different models of
source brightness profile that correspond to the magnification pattern
of Fig.~\ref{simplecau}} \label{simple_curves}
\end{figure}

 These results can depend on a complexity of
the caustic involved into our consideration (i.e., there can be
parts of the fold caustic close to the cusp points, or there can be
dense aggregations of caustics). From many-year observations of the
light curves of such GLS as Q2237+030, one can rule out such complex
cases. Therefore, we considered some modification of our statistical
consideration with rather simple fold caustic crossings.
However, the numerical results of this modification with simple
caustic crossing events (such as in
Fig.~\ref{simplecau},\ref{simple_curves} as an example) appeared to
be nearly the same as those of Table \ref{differences1}. As an
example, Table \ref{differences4} shows the results of
simulations analogous to those in Table \ref{differences1} with the r.m.s.
radius $R_{\rm rms}=0.21$; here, we excluded the models with $R_{\rm
rms}=\infty$ (here, $p=3$ for the PL model). The larger error is due
to a smaller number of realizations with the ``simple''
caustic.\looseness=1

\subsection{Gaussian fittings of the accretion disk and limb darkening models}
\label{GaussianFitting}

The above results concern with a comparison of different models
\textit{with the same} $R_{1/2}$ or $R_{\rm rms}$. However, in
reality, we do not know the source model to fit, and one may ask why
a light curve is not fitted within a different model. Therefore, one
must verify whether we can replace one model with a different one
with some other source parameters to get a better fitting.

\begin{table}[b]
\noindent\caption{Differences between source's models with the same
\boldmath$R_{\rm rms}=0.21 R_{\rm E}$ parameter in simple caustic
crossing events; here, $p=3,\, q=1$}\vskip0.5mm \tabcolsep18.0pt

\noindent{\footnotesize \label{differences4}
\begin{tabular}{c c c}
 \hline%
 \multicolumn{1}{c}{\rule{0pt}{9pt}$i$-th model}%
 & \multicolumn{1}{|c}{$j$-th model}
 & \multicolumn{1}{|c}{\bf $\eta$ }\\%
\hline%
\rule{0pt}{9pt}{ AD1 } & { Gaussian } & { $0.12 \pm 0.04 $ }\\
{ LD  } & { Gaussian } & { $0.03 \pm 0.01$ } \\
{ PL   } & { Gaussian } & { $0.05 \pm 0.012$ } \\
{ AD1   } & { LD } & { $0.14 \pm 0.05$ } \\
{ AD1   } & { Power-law } & { $0.08 \pm 0.03$ } \\
{ LD   } & { PL } & { $0.08 \pm 0.02$ } \\
\hline
\end{tabular}
}
\end{table}

We have fitted the limb-darkening and accretion disk model light
curves with that of the Gaussian source of different radii. The
half-brightness radius varied from $R_{1/2}=0.2 R_{\rm E}$ to
$R_{1/2}=0.24 R_{\rm E}$; here, $p=3/2$ and $q=1$.

\begin{figure}
\includegraphics[width=\column]{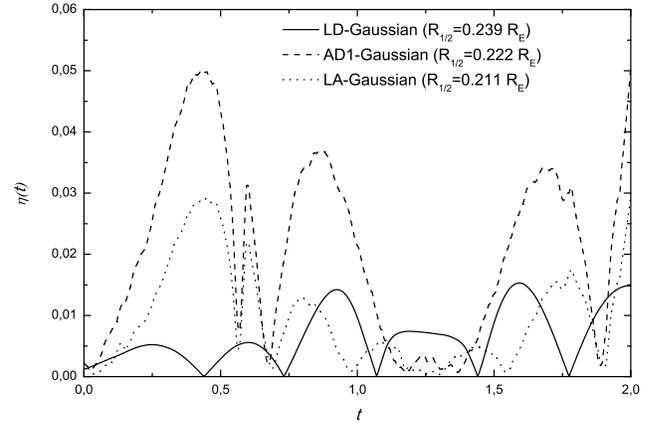}
\vskip-3mm\caption{Relative differences between the Gaussian, LD,
AD1, and LA models for best fitted curves} \label{relat_diffs}
\end{figure}

As we can see from Fig.~\ref{relat_diffs} and Table~\ref{differences3}, the Gaussian source cannot
reproduce all the models, though the fitting results are rather good for the class of compact
models. For example, the LD model can
be replaced by the Gaussian model with the other source size. The
relative differences between models in Fig.~\ref{relat_diffs} and Table~\ref{differences3} show
that it is possible to fit a considerable part of the whole curve
with different source models on the accuracy level which is
comparable to that of modern photometric observations.

\begin{figure}
\includegraphics[width=\column]{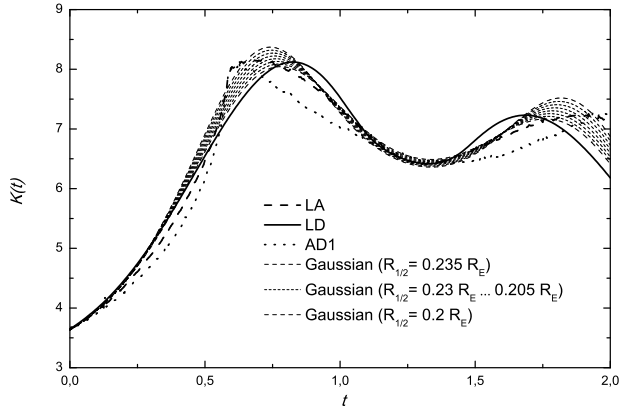}
\vskip-3mm\caption{Light curves of the LD, AD1 and AL models are fitted
with a Gaussian source of different sizes from $R_{1/2}=0.2
R_{\rm E}$ to $R_{1/2}=0.24 R_{\rm E}$ with step $\delta R_{1/2} =
0.001 R_{\rm E}$. These light curves correspond to the magnification
pattern of Fig.~\ref{simplecau}} \label{simple_fitt_curves}
\end{figure}

\begin{table}[b]
\noindent\caption{Differences between models for fitted
curves}\vskip3mm\tabcolsep8.8pt

\noindent{\footnotesize \label{differences3}
\begin{tabular}{c c c}
 \hline%
 \multicolumn{1}{c}{\rule{0pt}{9pt}$i$-th model}%
 & \multicolumn{1}{|c}{$j$-th model}
 & \multicolumn{1}{|c}{\bf $\eta$ }\\%
\hline%
\rule{0pt}{9pt}{ AD  } & { Gaussian ($R_{1/2}=0.2$) } & { $0.07 \pm 0.006 $ }\\
{ LD  } & { Gaussian ($R_{1/2}=0.24$) } & { $0.026 \pm 0.002$ } \\
{ LA   } & { Gaussian ($R_{1/2}=0.2$) } & { $0.034 \pm 0.006$ } \\
\hline
\end{tabular}
}
\end{table}

\section{Approximation Methods}
\label{s2}

\subsection{Statement of the problem}
\label{ss2.1}

The study of a caustic crossing event is closely related to
the investigation of the lens equation solutions.  This equation near a
fold can be expanded in powers of local coordinates; in the lowest
orders of this expansion, the caustic is represented by a straight
line; so, this approximation is often referred as  a ``linear
caustic approximation''. In this approximation, the point source
flux amplification is given by a simple formula (\ref{flux_extended
linca}) including the distance to the caustic and
two fitting parameters. In the most known cases, the linear caustic
approximation is sufficient to fit the observed light curves during
HAEs at the modern accuracy of flux measurements. At the same time,
the consideration of ``post-linear'' terms is sometimes necessary to
explain the present observational data. The need for a modification of
this formula -- e.g., by taking the caustic curvature into account
-- is being discussed for a long time
\citep{Fluke,shalyapin_01,Congdon_08}. The corrections to the
amplification coefficient in the case of the macrolensing were the
subject of investigations dealing with the problem  of ``anomalous
flux ratios'' \citep*{keeton_05}. A modification of the post-linear
approximation allows one to improve the quality of the fitting of the HAE
light curve of the image C in the GLS ``Einstein Cross''
\citep{alzhf_10,alzh_10}. One may hope for that an improvement of the
photometric accuracy of GLS observations will make it possible to
obtain additional parameters of the lens mapping connected with the
mass distribution in the lensing galaxy. Below, we study some points
of an approximate solution of the lens equation near the fold caustic
which probably is crossed by the source in many HAEs.

The standard consideration of the caustic crossing events deals
with the Taylor expansion of the potential near some point
 of the critical curve in the image plane. Let this point
be the coordinate origin in an appropriate coordinate system, and
let this point map onto the  coordinate origin of the source plane.
Further, we rotate the coordinate systems in both planes so that
the abscissa axis on the source plane
is tangent to the caustic at the origin, the quantity $|y_2|$
defines locally the distance to the caustic, and $y_1$ defines a
displacement along the tangent. For the harmonic potential,
we can write the  corresponding lens equation as
\begin{equation}
\label{eq3}
\begin{array}{l}
 y_1 = 2x_1 + a\left( {x_1 ^2 - x_2 ^2} \right) + 2bx_1 x_2 + c\left( {x_1 ^3 - 3x_1 x_2 ^2} \right) - \vspace{2\jot}\\
  - d\left( {x_2 ^3 - 3x_2
x_1 ^2} \right) + gx_2 ^4 + ..., \vspace{3\jot}\\
 y_2 = \mbox{ } b\left( {x_1 ^2 - x_2 ^2} \right) - 2ax_1 x_2 +  d\left( {x_1 ^3 - 3x_1 x_2 ^2} \right) + \vspace{2\jot}\\
  +c\left( {x_2 ^3 - 3x_2
x_1 ^2} \right) + fx_2 ^4 + ...,
\end{array}
\end{equation}
\noindent where $a$, $b$, $c$, $d$, $g$, and $f$ are expansion
coefficients. If the $y_2$ axis is directed toward the convexity of
the caustic, then $b<0$ (at fold points, $b\neq0$).

We now proceed to the derivation of approximate solutions of
Eqs.~(\ref{eq3}). To do this, we present two different methods
\citep{alzh_10}. The first method deals with analytic expansions
in powers of a small parameter. However, it results in
nonanalytic functions of coordinates leading to nonintegrable
terms in the amplification coefficient. The second method does not
lead to such problems, though it uses a somewhat more complicated
representation of the solution of the lens equation containing
square roots of analytic functions. The methods agree with each
other in a common domain of validity; moreover, we use the second
method to justify some expressions in the amplification formulas
in terms of distributions to validate the applications to extended
source models. A more detailed presentation may be found in
\citep{alzh_10}.

First, we suppose that the source and the caustic lie on different
sides from the $y_1$ axis. Then, $y_2
> 0,$ and we substitute
\begin{equation}
\label{eq4}
 y_i = t ^ {2} \y_i, \quad x_1 = t ^ 2 \x_1, \quad x_2 = t \x_2,
\end{equation}
\noindent where $i=1,2,$ and $t$ can be considered as a parameter
of vicinity to the caustic. This is a formal substitution that
makes easier the operations with different orders of the expansion.
After calculations, we put $t=1$ and thus return
to the initial variables $y_i$. However, if we put $\y_i$ to be
constant with varying $t$, then this substitution allows us to
study a local behavior of critical image trajectories; $t=0$
corresponds to crossing the caustic by a point source, when two
critical images appear. As was shown in \cite{alzh_10}, this allows one to look for
solutions of Eqs.~(\ref{eq3}), by using the expansions of $\x_i$ in powers of $t$:
\[
\x_1 = \x_{10} + \x_{11} t + \x_{12} t^2+...,
\]
\begin{equation}
\label{eq5}
 \x_2 = \x_{20} + \x_{21} t + \x_{22} t^2+....
\end{equation}
It should be stressed that the analyticity in $t$ does not mean that
the coefficients of expansions~(\ref{eq5}) will be analytic
functions of coordinates $\y_i$ in the source plane (see below).

In terms of the new variables~(\ref{eq4}), system~(\ref{eq3}) takes
the form (up to the terms $\sim t^2$)
\[
 \y_1 = 2\x_1 - a\x_2^2 + t(2b\x_1 \x_2 - d\x_2^3  )+\]
 \[
+ t^2( a\x_1^2 - 3c\x_1 \x_2^2 + g\x_2^4 ),
 \]
 \[
\y_2 =  - b\x_2^2 + t(  - 2a\x_1 \x_2 + c\x_2^3)+
 \]
\begin{equation}
\label{eq25}
 + t^2( b\x_1^2 - 3d\x_1 \x_2^2 + f\x_2^4 ).
 \end{equation}
The substitution of expansions~(\ref{eq5})
into~(\ref{eq25}) allows us to determine all
coefficients successively. For example, for the zero-order terms,
we have
\begin{equation}\label{eq6}
\x_{10} =
\frac{1}{2}\left( {\y_1 - \frac{a}{b}\y{ }_2} \right), \quad
\x_{20} = \varepsilon \sqrt { {\y_2 } \mathord{\left/ {\vphantom
{{\y_2 } b}} \right. \kern-\nulldelimiterspace}
\left|b\,\right|}\;,
\end{equation}
\noindent where $\varepsilon =\pm 1 $ determines two different
critical solutions. This approximation yields the well-known
formula for the amplification (\ref{flux_extended linca}).

The first-order approximation  terms
 have been derived in \citep{alzh_03,keeton_05}.
In microlensing observations, two critical images cannot be
resolved, so we need the total amplification coefficient of two
critical images.  However, the contributions of the  order of
$\sim t$ appear to be cancelled in calculations of the total
amplification. Therefore, to obtain a nontrivial correction to the
zero-order amplification, higher order approximations should be
involved. These corrections have been derived in
\cite{alzhf_10,alzh_10}. Note that the corresponding second-order
terms contain expressions nonanalytic in $\y_2$. Though the
second-order corrections are expected to be small, they appear to
be noticeable in some cases even in the analysis of the available
data on light curves in the Q2237+0305 GLS.

\subsection{Method 2}
\label{ss2.2}

The other approach to the construction of approximate solutions of
the lens equation in a vicinity of the fold is described in
\cite{alzh_10}. This allows us to present the critical solutions
of system~(\ref{eq25}) in the following form:
\[
\x_1 = p + tr\varepsilon \sqrt w ,
\]
\begin{equation} \label{new1}
 \x_2 = t\bar s + \varepsilon \sqrt w ,\quad
\varepsilon = \pm 1.
\end{equation}
After the substitution of
Eqs.~(\ref{new1}) in the lens equation, the
terms containing integer and half-integer powers of $w$ can be
separated \cite{alzh_10}. After some algebra, this yields a
system that is convenient for an iterative procedure resulting in
analytic expansions for the functions $p,r,\bar s,w $ both in
powers of $t$ and $\y_i$ at every step of iteration
\cite{alzh_10}. In \cite{alzh_10}, such a solution was obtained up
to the terms $\sim t^2$. If we expand $\sqrt w$ in powers of $t$,
then we immediately have the solution in the form~(\ref{eq5}).

\subsection{Amplification of a point source}
\label{ss3.1}

The solutions of the lens equation are then used to derive the
Jacobians of the lens mapping (for both images near the critical
curve). According to (\ref{eq2}), the value of $J^{ - 1}$ yields
the amplification of individual images. As we pointed out above,
we need the total amplification of two critical images (the sum of
two amplifications of the separate critical images). In the second-order
approximation (using the expansion up to the terms $~t^2$),
this is
\begin{equation} \label{point amplification} K_{\rm cr} = \frac{1}{2}\frac{\Theta
\left( {y_2 } \right)}{\sqrt {\left| b \right|y_2 } }\left[ {1 +
Py_2 + Qy_1 - \frac{\kappa }{4}\frac{y_1^2 }{y_2 }\mbox{ }}
\right], \end{equation}
\noindent where the constants $P$ and $Q$ are
expressed via the Taylor expansion coefficients from
Eq.(\ref{eq1}), and
\[
\kappa = {\displaystyle \frac{a^2 + b^2}{2\left| b \right|}} \, ;
\]
$\Theta {(y_2) }$ is the Heaviside step function. Note that $\kappa$
is the caustic curvature at the origin which enters explicitly into
the amplification formula. The parameters $P$ and $Q$ are independent;
the explicit formulae for them
can be found in \citep{alzhf_10,alzh_10}. However,
this is not needed when we use Eq.(\ref{point amplification}) for
fitting the observational data, because these constants are
whatever considered as free fitting parameters.

Formula~(\ref{point amplification}) yields an effective
approximation for the point source amplification near the coordinate
origin provided that $y_2> 0,$ and ${y_2}/{y_1^2}$ is not too small
(see the term containing $\kappa$). For a fixed source position, this
can be satisfied always by an appropriate choice of the coordinate
origin, so that the source will be situated almost on a normal
to the tangent to the caustic.

If the source is on the caustic tangent or in the region between
the caustic and the tangent, then formula~(\ref{point
amplification}) does not give a good approximation to the
point source amplification. Nevertheless, in the case of an extended
source, we will show that
result~(\ref{point amplification}) can be used to obtain
approximations to the amplification of this source even as it
intersects the caustic. However, to do this, we need to redefine
correctly the convolution of (\ref{point amplification}) with a
brightness distribution.

\section{Amplification of Extended Sources}
\label{s3}
\subsection{Transition to extended sources}
\label{ss3.2}

We now present the results of studies of the point source
amplification within extended source models. Let $I({\mathbf y})$ be a
surface brightness distribution of an extended source. If the
source center is located at the point ${\mathbf{Y}}=(Y_1,Y_2)$ in
the source plane, then the total microlensed flux from the source
is given by (\ref{flux_extended}).

Formula~(\ref{point amplification}) contains the nonintegrable term
$\sim\Theta(y_{2})  ( {y_2 })^{ - 3 /2} $. Therefore, the question
arises of how formula~(\ref{point amplification}) can be used in
the situation where the extended source intersects a caustic and some
part of the source is in the zone between the tangent and the
caustic. In view of Section~\ref{s2}, it is evident that the
mentioned term is a result of the asymptotic expansion of the root
$\sqrt {y_2 + \kappa y_1^2 t^2/2+...}$  in the approximate solution
(\ref{new1}). The direct usage of a solution in the form  (\ref{new1})
for the calculation of the Jacobians of the lens mapping and then for
the derivation of amplifications does not lead to any divergences,
and any nonintegrable terms in $K_{\rm cr}$ do not arise.
Nevertheless, it is convenient to have a
representation of $K_{\rm cr}$ in the form of an expansion in powers of
a small parameter. Such an expansion can be carried out correctly after
the substitution of $K_{\rm cr}$ into integral (\ref{flux_extended}). On
this way, starting from (\ref{new1}), it is easy to show that,
to define $K_{\rm cr}$ correctly, one must replace the term
$\Theta(y_2) ( y_2 )^{ - 3 /2} $ in (\ref{point amplification}) by
the distribution (generalized function) $( y_2 )_+^{ - 3/2}$
\citep{Gel'fand_64}. We recall that the distribution $ y _+^{ -
3/2}$ of the variable $y$ is defined by the expression
\[
\int{ y _ + ^{-3 / 2} f( y)dy} =  2\int\limits_0^\infty {y^{- 1/
2}} \frac{\partial f( y)}{\partial y}dy
\]
\* for any test function $f(y)$.

After this redefinition, we have
\begin{equation}
\label{generalized Kcr} K_{\rm cr} = {\displaystyle \frac{\Theta
\left( {y_2 } \right)}{2\sqrt {\left| b \right|y_2 } }}\left[ {1 +
Py_2 + Qy_1 \mbox{ }} \right] - {\displaystyle \frac{\kappa }{8\sqrt
{\left| b \right|} }}{y_1^2 }{\left( y_2 \right)_ + ^{-3 / 2} }.
\end{equation} \*This formula can be used to correctly derive an
approximate amplification of a sufficiently smooth extended source
including the case where the source crosses the caustic.

\subsection{Gaussian source}
\label{ss3.3}

Formula~(\ref{generalized Kcr}) has been used \citep{alzh_10} to
derive the amplification of a Gaussian source with the brightness
distribution (\ref{Gauss}), the limb-darkening source, and the
power-law source (see the next subsections).

Further, we use the dimensionless coordinates $s = Y_1/R,\, h =
Y_2/R$ of the source center and the functions
\[
I_k \left( h \right) = \int\limits_0^\infty {u^{k - 1/ 2}\exp \left(
{ - u^2 + 2uh} \right)du} =
\]
\begin{equation}
\label{eq16} = \frac{1}{2}\sum\limits_{n = 0}^\infty {{\displaystyle
\frac{\Gamma \left( {\textstyle{1 \over 4} + \textstyle{{k + n}
\over 2}} \right)}{n!}}} ( {2h})^n.
\end{equation}
These functions can be expressed in terms of the confluent
hypergeometric function  $_1F_1 $ or the parabolic cylinder
function $D$:
\begin{equation}
\label{eq17}
 I_k \left( h \right) =
 2^{ - \left( {\frac{k}{2} + \frac{1}{4}} \right)}\Gamma \left(
{k + \frac{1}{2}} \right)e^{\frac{h^2}{2}}D_{ - \left( {k +
\frac{1}{2}} \right)} \left( { - \sqrt 2  h} \right).
\end{equation}

The substitution of (\ref{generalized Kcr})  and (\ref{Gauss})
in (\ref{flux_extended}) yields
\[
 K_G ( {s,h})= {\displaystyle \frac{1}{2\sqrt {\pi \left| b
\right|R} }} \left\{ \vphantom {\left [{\displaystyle \frac{\kappa
}{2}}\right]} \Phi_0 \left( h \right)+ \right.
\]
\begin{equation}
\label{eq18}
  + R\left[ \vphantom
{{\displaystyle \frac{\kappa }{2}}}P \Phi_1 \left( h \right) -
\right.
 {\displaystyle \frac{\kappa }{2}} \Phi_2\left( h \right) + Q  s  \Phi_0 \left( h \right) \left.
{\left. {- \;\vphantom {{\displaystyle \frac{\kappa }{2}}}\kappa
s^2 \Phi_2}\left( h \right) \right]} \right\} \mbox{ }.
\end{equation}
\noindent
Here,
\[
 \Phi _0 \left( h \right) = I_0 \left( h \right)\exp \left( { - h^2} \right),
\]
\[ \Phi _1 \left( h \right) = I_1 \left( h \right)\exp \left( { -
h^2} \right),
\]
\[
\Phi _2 \left( h \right) = \left[ {hI_0 \left( h
\right) - I_1 \left( h \right)} \right]\exp \left( { - h^2}
\right).
 \]
\* Note that the main term of (\ref{eq18}) which corresponds to
the linear caustic approximation was first obtained in work
\cite{Schneider Weiss}.

\subsection[CLD]{Limb-darkening source}
\label{CLD}

Analogous considerations allow us to obtain formulas for the
amplification of extended sources for the limb-darkening and
power-law brightness profiles; the results are represented
analytically in terms of the hypergeometric function $_2F_1$
\citep{bateman}.

Denote
\[
{\rm X}_{k,q} \left( h \right) = \frac{\Gamma \left( {q + 2}
\right)}{\Gamma \left( {q + \frac{3}{2}}
\right)}\int\limits_0^\infty {y^{k - \frac{1}{2}}\,\,\Xi( y -
h;q+1/2)\,dy},
\]
\* $k=1,2$. We have
\[
X_{k,q}( h) = 2^{q + \frac{1}{2}}(1 + h)^{q + k +
1}\frac{\Gamma(q+2)\Gamma(k + \frac{1}{2})} {\Gamma(q + k + 2)}\,
\times
\]
\[\times \,\,\, { }_2F_1 \left( { - q - \frac{1}{2},q + \frac{3}{2};q
+ k + 2;\frac{1 + h}{2}} \right)
\]
for $ - 1 < h < 1$ and
\[
{\rm X}_{k,q} \left( h \right)=
\]
\[ = \sqrt \pi \left( {h + 1}
\right)^{k - \frac{1}{2}}{ }_2F_1 \left( {q +
\frac{3}{2},\frac{1}{2} - k;2q + 3;\frac{2}{h + 1}} \right)
\]
for $h > 1.$

For $k=-1,$  we define
\[{\rm X}_{ - 1,q}
\left( h \right) = 4\left( {q + 1} \right)\left( {h{\rm X}_{0,q - 1}
- {\rm X}_{1,q - 1} } \right).
\]

Then, in case of the model with limb darkening (\ref{C2}), the
critical images disappear when the source lies on the outer side
of the caustic (i.e., for $h < - 1$). The amplification due to
critical images takes the form
\[
 K_{\rm LD} ( {s,h} ) =
\frac{1}{2\sqrt {\pi | b|R} } \Big\{ {{\rm X}_{0,q}(h) + R\Big[{
\vphantom {\Big [{\displaystyle \frac{\kappa }{2}}\Big]} P{\rm
X}_{1,q}(h)-} }
\]
\[
{{ - \frac{\kappa}{8 ( q + 2)}{\rm X}_{ - 1,q + 1}(h) + Q\,s\,{\rm
X}_{0,q} (h) - \frac{\kappa }{4}\,s^2{\rm X}_{ - 1,q} (h) }\Big]}
\Big\}. \]

\subsection[C]{Amplification for a power-law source}
\label{C}
The result for the amplification involves integrals:
\[
\Psi _{k,p} \left( h \right) = \frac{\Gamma \left( {p -
\frac{1}{2}} \right)}{\Gamma \left( {p - 1} \right)}
\int\limits_0^\infty {\frac{y^{k - \frac{1}{2}}dy}{\left( {1 +
\left( {y - h} \right)^2} \right)^{p - 1 / 2}}}=
\]
\[
 = \frac{\Gamma
\left( {p - \frac{1}{2}} \right)}{\Gamma \left( {p - 1} \right)}
B\left( {k + \frac{1}{2},2p - k - \frac{3}{2}} \right)\, (1 +
h^2)^{k/2+3/4-p} \,\times \]
\[
\times\,\, { }_2F_1 \left( {k + \frac{1}{2},2p - k -
\frac{3}{2};\,p\,\,;\frac{1}{2}\left( {1 + \frac{h}{\sqrt {1 +
h^2} }} \right)} \right)
\]

\*for $k = 0,1$, $B\left( {x,y} \right)$ being the Beta-function.

We extend this to $k = -1$ in view of the definition of
$(y)_+^{ - 3/2}.$ So, we have
\[
\Psi _{-1,p} ( h) = 4(p-1)[ h\Psi _{0,p+1} ( h) -
\Psi _{1,p+1} ( h)].
\]
\*Now, the amplification due to critical images takes the form
\[
 K_{\rm PL}( s,h ) =  \frac{1}{2\sqrt {\pi |b|R }}
 \Big\{ \Psi_{0,p}( h) + R \Big[ {
 \vphantom {\Big [{\displaystyle \frac{\kappa }{2}\Psi _{-1,p}}\Big]}P \Psi_{1,p}(h) -
 }   \]
\[   -  \frac{\kappa }{8(p-2) }\Psi _{-1,p-1}(h) + Q
\,s\, \Psi _{0,p}(h) - \frac{\kappa}{4}\, s^2\Psi _{-1,p}(h)  \Big]
\Big\}.  \] \*The zeroth approximation to this formula has been
derived in \cite{shalyapin_01}.

\section{Discussion}
\label{discussion}
 \noindent Prior to sum up the results of this
paper, we must stress that the consideration of HAEs without some
additional information may only provide a very uncertain, on
order-of-magnitude level, and even ambiguous information about the
source parameters. For example, one may expect that, by adding a
sufficient number of small microlenses, it is possible to reproduce
fine features of a light curve during a HAE and to fit the light
curve of any compact source. The introduction of planetary mass
objects in the lensing galaxy (which is quite reasonable, see, e.g.,
\cite{wyithe_01,tsapras_03,Mao_91,Jaroszynski_02}) yields a fine
caustic mesh, which, in turn, can give rise to fine features of
light curves during a HAE similar to those due to any given
brightness distribution over the source. Therefore, we point out the
assumptions and the class of models involved, which is necessary for
a reasonable formulation of the problem.\looseness=1

The simulations of the present paper use the GLS parameters similar
to those of Q2237+030 \cite{wyithe_01, Schmidt_98}. We consider the
equal mass microlensing system; we do not consider any mass
distributions and/or populations of small (planetary) masses. Next,
we consider the most simple source models without effects of
ellipticity, {\it etc.} Nevertheless, we see that, even under these
rather severe restrictions, there is an ambiguity in the determination
of the source model from the observations of HAEs.  The limb-darkening
source can be successively fitted with the Gaussian source model
(with the other $R_{1/2}$) on 1-2\% accuracy level. This confirms
the conclusions of Mortonson {\it et al.} \cite{Mortonson_05} (see also
\cite{Congdon_07}) that the surface brightness profile has little
effect on microlensing. What can be determined within the modern
accuracy is the source size; one can also distinguish either we deal
with a compact source or not (i.e., with a steep decrease of the
brightness for large radii).

The investigation of the source structure from HAEs is closely
related to the solution of the lens equation in the caustic region.
There are also different approaches to this problem under different
restrictions. In this paper, we confined ourselves to the case of the
fold caustic. We outlined two methods that enable us to obtain the
critical solutions of the gravitational lens equation near a fold
with any desired accuracy \cite{alzh_10,alzhf_10}. In order to
obtain nontrivial corrections to $K_{\rm cr}$ obtained in the
linear caustic approximation, the higher orders of the expansion of
the lens equation must be taken into account, as compared to works
\citep{alzh_03,keeton_05}. The modified formula for $K_{\rm cr}$
contains three extra parameters. This is applied to the Gaussian,
power-law, and limb darkening models of an extended source.  The
fitting of the light curve of GLS Q2237+0305C using these modified
relations shows \cite{alzh_10,alzhf_10} that some of these
corrections can be statistically significant even at the present
accuracy level. This means that if we are looking for some fine
effects in HAEs due to the source size, a consistent treatment must
involve sometimes the higher-order corrections to
solutions of the lens equation.

\vskip3mm This work has been supported in part by the
``Cosmomicrophysics'' program of the National Academy of Sciences of
Ukraine. We are grateful to the staff of the Computer Center of
National Technical University of Ukraine ``Kyiv Polytechnic
Institute'', where a considerable part of our simulations has been
performed.

\rezume{%
ЕФЕКТИ ПЕРЕТИНУ КАУСТИКИ І МОДЕЛІ ДЖЕРЕЛА\\ У ГРАВІТАЦІЙНО-ЛІНЗОВИХ
СИСТЕМАХ}{О.М. Александров, В.М. Слюсар, В.І. Жданов}{Події з
великим підсиленням (ПВП) є звичайним явищем у позагалактичних
гравітаційно-лінзових системах (ГЛС), де спостерігають декілька
зображень віддаленого квазара на фоні галактики, що знаходиться на
передньому плані. Протягом ПВП відбувається значне збільшення
яскравості в одному із зображень квазара. Грігер, Кайзер та Рефсдал
запропонували використовувати ПВП для вивчення центральних областей
квазара в ГЛС. У цій статті  досліджуємо пов’язані з цим питання, що
стосуються ідентифікації різних типів джерела на базі спостережень
ПВП. Ми порівнюємо результати числових моделювань кривих блиску для
того, щоб оцінити можливість відрізнити моделі джерела в ГЛС.
Запропоновано наближені схеми для розв’язання рівняння лінзи в околі
каустики -- складки. Результати використано для отримання
коефіцієнтів підсилення, що враховують поправки високого порядку для
моделей гаусівського і степеневого джерела, а також джерела з
потемнінням до краю. }


\begin{thebibliography}{9} %

\bibitem{schneider_92}
P.~Schneider, J.~Ehlers, and E.E.~Falko, {\it Gravitational Lenses}
(Springer, New York, 1992).
\bibitem{Wambsganss_06}
J.~Wambsganss, {\it Gravitational Lensing: Strong, Weak, and Micro}, edited by
G.~Meylan, P.~North, and P.~Jetzer (Springer, Berlin, 2006),
p.~453.
\bibitem{grieger_88}
B.~Grieger, R.~Kayser, and S.~Refsdal, A\&A \textbf{194}, 54 (1988).
\bibitem{Huchra_85}
J.~Huchra, V.~Gorenstein, S.~Kent, I.~Shapiro, G.~Smith, E.~Horine,
and R.~Perley, AJ \textbf{90}, 691 (1985).
\bibitem{alcald_02}
D.~Alcalde, E.~Mediavilla, O.~Moreau, J.A.~Munoz,  C.~Libbrecht {\it
et al.}, ApJ \textbf{572}, 729 (2002).
\bibitem{udalski_2006}
A.~Udalski {\it et al.}, Acta Astron. \textbf{56}, 293 (2006).
\bibitem{wozniak_00}
P.R.~Woz\'niak, C.~Alard, A.~Udalski, M.~Szyma\'nski, M.~Kubiak,
G.~Pietrzy\'nski, and K.~Zebru\'n,  ApJ \textbf{529}, 88 (2000).
\bibitem{Bogdanov_02}
M.B.~Bogdanov and A.M.~Cherepashchuk, Astron. Repts.  \textbf{46},
626 (2002).
\bibitem{shalyapin_01}
V.N.~Shalyapin, Astron. Lett.  \textbf{27}, 150 (2001).
\bibitem{shalyapin_02}
V.N.~Shalyapin, L.J.~Goicoechea, D.~Alcalde, E.~Mediavilla,
J.A.~Mu\~noz, and R.~Gil-Merino, ApJ \textbf{579}, 127 (2002).
\bibitem{wyithe_99}
J.S.~Wyithe, R.L.~Webster, and E.L.~Turner, MNRAS \textbf{309}, 261
(1999).
\bibitem{wyithe_00b}
J.S.~Wyithe, R.L.~Webster, and E.L.~Turner, MNRAS  \textbf{318}, 762
(2000).
\bibitem{wyithe_00a}
J.S.~Wyithe, R.L.~Webster, E.L.~Turner, and D.J.~Mortlock, MNRAS
\textbf{315}, 62 (2000).
\bibitem{yonehara_01}
A.~Yonehara, AJ  \textbf{548}, 127 (2001).
\bibitem{gil_06}
R.~Gil-Merino, J.~Gonzalez-Cadelo, L.J.~Goicoechea, V.N.~Shalyapin,
and G.F.~Lewis, MNRAS \textbf{371}, 1478 (2006).
\bibitem{goico_01}
L.J.~Goicoechea, D.~Alcalde, E.~Mediavilla, and  J.A.~Mu\~noz, A\& A
\textbf{397}, 517 (2003).
\bibitem{Mortonson_05}
M.J.~Mortonson, P.L.~Schechter, and J.~Wambsganss, ApJ
\textbf{628}, 594 (2005).
\bibitem{Kochanek_04}
C.S.~Kochanek, ApJ \textbf{605}, 58 (2004).
\bibitem{anguita_08}
T.~Anguita, R.W.~Schmidt, E.L.~Turner, J.~Wambsganss, R.L.~Webster,
K.A.~Loomis, D.~Long, and R.~MacMillan, A\&A \textbf{480}, 327
(2008).
\bibitem{Poindexter_10a}
S.~Poindexter and C.S.~Kochanek, ApJ \textbf{712}, 658 (2010).
\bibitem{Poindexter_10b}
S.~Poindexter and C.S.~Kochanek, ApJ \textbf{712}, 668 (2010).
\bibitem{Poindexter_08}
S.~Poindexter, N.~Morgan, and C.S.~Kochanek, ApJ  \textbf{673}, 34
(2008).
\bibitem{vakulik_07}
V.G.~Vakulik, R.E.~Schild, G.V.~Smirnov, V.N.~Dudinov, and
V.S.~Tsvetkova, MNRAS \textbf{382}, 819 (2007).
\bibitem{dominik}
M.~Dominik, MNRAS  \textbf{353}, 69 (2004).
\bibitem{schakura_73}
N.I.~Shakura and R.A.~Sunyaev, Astron. Astrophys. \textbf{24}, 337
(1973).
\bibitem{Schmidt_98}
R.~Schmidt, R.L.~Webster, and F.G.~Lewis, MNRAS \textbf{295}, 488
(1998).
\bibitem{Dai_03}
X.~Dai, E.~Agol, M.W.~Bautz, and G.P.~Garmire, ApJ \textbf{589}, 100
(2003).
\bibitem{Fedorova_09}
E.V.~Fedorova, V.I.~Zhdanov, C.~Vignali, and G.G.C.~Palumbo, A\&A
\textbf{490}, 989 (2008).
\bibitem{Vakulik_06}
V.~Vakulik, R.~Schild, V.~Dudinov, S.~Nuritdinov, V.~Tsvetkova,
O.~Burkhonov, and T.~Akhunov, A\&A \textbf{447}, 905 (2006).
\bibitem{Fluke}
C.J.~Fluke and R.L.~Webster, MNRAS  \textbf{302}, 68 (1999).
\bibitem{Congdon_08}
A.B.~Congdon, C.R.~Keeton, and C.E.~Nordgren, MNRAS \textbf{389},
398, (2008).
\bibitem{keeton_05}
C.R.~Keeton, B.S.~Gaudi, and A.O.~Petters, ApJ \textbf{635}, 35
(2005).
\bibitem{alzh_10}
A.N.~Alexandrov and V.I.~Zhdanov, e-print arXiv:1006.5903 (2010).
\bibitem{alzhf_10}
A.N~Alexandrov, V.I~Zhdanov, and E.V.~Fedorova, Astron. Lett.
\textbf{36}, 329  (2010).
\bibitem{alzh_03}
A.N~Alexandrov, V.I~Zhdanov, and E.V.~Fedorova, Visn. Kyiv.
Univ., Astron. \textbf{39-40}, 52 (2003).
\bibitem{Gel'fand_64}
I.M.~Gel'fand and G.E.~Shilov, {\it Generalized Functions, Vol. 1}
(Academic Press, New York, 1964).
\bibitem{Schneider Weiss}
P.~Schneider and A.~Weiss, A\&A  \textbf{171}, 49 (1987).
\bibitem{bateman}
H.~Bateman and A.~Erd\'elyi, {\it Higher Transcendental Functions,
Vol. 1} (McGraw-Hill, New York, 1953).
\bibitem{Jaroszynski_02}
M.~Jaroszynski, B.~Paczynski, AA \textbf{52}, 361 (2002).
\bibitem{Mao_91}
S.~Mao and B.~Paczynski, ApJ \textbf{374}, 37 (1991).
\bibitem{tsapras_03}
Y.~Tsapras, K.~Horne, S.~Kane, and K.~Carson, MNRAS  \textbf{343},
1131 (2003).
\bibitem{wyithe_01}
J.S.B.~Wyithe and E.L.~Turner,  MNRAS \textbf{320}, 21 (2001).
\bibitem{Congdon_07}
A.B.~Congdon, C.R.~Keeton, and S.J.~Osmer, MNRAS \textbf{376}, 263
(2007).

\begin{flushright}
{\footnotesize Received 30.01.11}
\end{flushright}
\end{thebibliography}
\end{document}